\documentclass[%
 preprint,
 amsmath,amssymb,
 aps, physrev,
]{revtex4-2}

\usepackage{graphicx}
\usepackage{dcolumn}
\usepackage{bm}
\usepackage{hyperref}
\usepackage{academicons}
\usepackage{xcolor}
\usepackage{orcidlink}
\usepackage{lipsum}
 \usepackage{natbib}

\usepackage[nameinlink,noabbrev]{cleveref}
\crefname{equation}{Eq.}{Eqs.}
\Crefname{equation}{Equation}{Equations}
\crefname{figure}{Fig.}{Figs.}
\Crefname{figure}{Figure}{Figures}

\usepackage{siunitx}

\begin{document}

\preprint{APS/123-QED}

\title{\textbf{Peristaltic pumping under poroelastic confinement} 
}%

\author{Avery Trevino\orcidlink{0000-0003-3481-1437}}
\author{Roberto Zenit\orcidlink{0000-0002-2717-4954}}
\author{Mauro Rodriguez Jr.\orcidlink{0000-0003-0545-0265}}\email{Contact author: mauro\_rodriguez@brown.edu}
\affiliation{%
School of Engineering, Brown University, Providence, RI
}%

\date{\today}

\begin{abstract}
Low Reynolds number flow near a poroelastic interface can be found across scales in biological and engineered systems. 
We develop a 2D model of peristaltic flow confined under a poroelastic solid. 
In this geometry, the lower boundary is an infinite train of traveling waves which pump fluid along a channel. 
The upper boundary of the flow is a poroelastic half space.
The flow and deformation are solved analytically by an asymptotic expansion in the peristaltic amplitude and depend nonlinearly on dimensionless poroelastic stiffness, permeability, and interfacial slip. 
We quantify the effect of material properties on the poroelastic fluid-structure interaction. 
Peristaltic flow through the channel is inhibited by poroelastic confinement owing to increased viscous dissipation across the interface and energy loss in deforming the elastic solid.
Permeability and slip interact with the  material stiffness to produce material dependent regimes of forward or backward interstitial flow within the poroelastic domain.
The maximum Darcy flow is found to occur at permeability values that optimize the elastic matrix interaction.

\end{abstract}


\maketitle

\section{Introduction\label{sec:intro}}
Peristaltic pumping produces fluid transport from periodic deformations traveling along a boundary \citep{Shapiro1969,Floryan2021,Jaffrin1971,Fung1968,Provost1994,Takagi2011,selverov2001}. 
It generates flow in engineering and biological systems, e.g., in mechanical pumps~\cite{Forouzandeh2021} and the ureter and glymphatic system ~\cite{Shapiro1969, Takagi2011,Romano2020,Bauerle2020,Hadaczek2006,Wang2011,Mestre2018,Trevino2025}.
In engineered systems, peristalsis is commonly driven by rollers or sliders that press and glide over the exterior of flexible tubes \citep{Forouzandeh2021,Floryan_2022}.
Biomechanical peristalsis can arise due to vessel tissue contractions~\citep{Bauerle2020,Weinberg1971} or by pressure driven deformation~\citep{Thomas2019,Yokoyama2021,Wang2011,Mestre2018,Romano2020,Hadaczek2006}.
In a similar problem, the flow produced by an infinitely long undulating sheet has been studied for microbiological swimming~\cite{Taylor1951,katz_propulsion_1974,Dias2013,Shaik2019,hewitt2017taylor,Jha2025,Reynolds1965,Iqbal2025,Leshansky2009,Blake1971,Fu2010,Takagi2024,Tchoufag2019}.
Recent developments in microfluidic devices which mimic biological flows have raised interest in peristaltic pumping under poroelastic confinement.
Many of these devices, and their biological counterparts, contain compliant, permeable boundaries for controlled fluid mixing or particle transport in applications such as drug delivery~\cite{huang2025oscillatory,zilberman2021}.
It is critical to model the fluid-poroelastic interaction to predict and design flow in engineered devices and to understand fundamental biomechanical processes. 

The peristaltic wave shape and motion (i.e., transverse and longitudinal) affects the resulting induced flow.
\citet{Shapiro1969} were the first to establish relationships between wave parameters and flow rates in the low Reynolds number, long-wavelength limit of peristaltic pumping.
The flow rate in this limit is proportional to the peristaltic amplitude squared.
\citet{Bauerle2020} showed that the network-forming slime mold \textit{Physarum polycephalum} optimally combines transverse peristaltic wave modes to maximize the transport of nutrients.
\citet{Floryan_2022} found that tilted transverse wave forms produce greater flow as opposed to single-mode sinusoidal peristaltic profiles.
Longitudinal wall motion naturally arises during the radial contraction of elastic tubes and should not be neglected in the flow analysis.
These oscillations induce backward flow, i.e., reflux,  in the Eulerian frame~\citep{katz_propulsion_1974,Reynolds1965,Blake1971,Shaik2017} and suppress it in the Lagrangian frame~\citep{Trevino2025,winn2025}.
Peristalsis often occurs in compliant confinement where oscillatory flow interacts with a soft boundary~\citep{kelley2021,Yokoyama2021,mudugamuwa2024periodic,kim2012human,huh2010reconstituting,huang2025oscillatory}.
This confinement reduces the efficiency of peristaltic pumping due to fluid kinetic energy loss to elastic storage in the boundary~\citep{Trevino2025}. 
As argued above, many biological systems have poroelastic boundaries. 
The fluid flow through these permeable structures has been argued to play a significant role in the flow characteristics and remain not well understood.

Fluid flow through porous media has been studied in the context of biomechanical~\citep{moeendarbary2013cytoplasm,franceschini2006brain,lai1991triphasic,swartz2007interstitial,maiti2011peristaltic,mow1980biphasic}, engineered~\citep{kim2012human,koroleva2016hydrogel}, and groundwater flows~\citep{darcy,biot1941general,jha2014coupled,yarushina2013rock,rutqvist2012geomechanics,bear1981mathematical}.
\citet{darcy} first described the motion of fluid through a porous solid in experimental studies of water flowing through sand.
Darcy's law states the mean velocity of interstitial fluid is proportional to the pressure gradient by the ratio of permeability to fluid viscosity.
These empirical observations were later derived theoretically by \citet{whitaker1986flow}.
\citet{brinkman1949calculation} added the Laplacian of the average velocity with an ``effective viscosity" coefficient as a viscous correction term which becomes important when viscous shear is non-negligible.
\citet{biot1941general} extended Darcy's law to include solid deformation in the theory of poroelastic consolidation. The theory couples the deformation-diffusion response of a biphasic material in which an interstitial fluid flows through the connected pores of an elastic solid skeleton.
Poroelastic consolidation theory has since been applied to geophysical flows~\citep{bear1981mathematical,rutqvist2012geomechanics,yarushina2013rock,jha2014coupled} and tissue mechanics~\citep{lai1991triphasic,franceschini2006brain,moeendarbary2013cytoplasm}.

More generally, the Stokes-Biot problem is a coupled system involving a poroelastic body interacting with viscous fluid~\citep{badia2009coupling, ruiz2022biot, Finney2024,Finney2025}. 
Exterior flow induces surface traction on the solid skeleton, generating pore pressure and internal flow which dissipate energy within the poroelastic body.
A key choice to study this problem is the interfacial boundary condition coupling the poroelastic body with the external flow. 
The empirical work of \citet{Beavers1967}, theoretically justified later by \citet{saffman1971boundary}, studied the rigid limit of this interaction.
The Beavers-Joseph-Saffman (BJS) boundary condition assumes Darcy flow within the porous region and the fluid velocity jump is proportional to the Stokes shear stress by a permeability length scale. 
\citet{ochoa1995momentum1} would later include the Brinkman correction to Darcy's law.
As a result, the interstitial velocity is proportional to a stress jump, instead of a velocity jump, across the porous interface.
\citet{Minale2014} suggested the Stokes fluid momentum be transferred to both the pore fluid and solid skeleton such that total traction is continuous.
The stress within the poroelastic medium is partitioned between the elastic solid and pore fluid weighted by their respective volume fractions.  
\citet{Feng2020} and \citet{xu} used a thermodynamic energy dissipation approach in which the interface velocity jump is proportional to the fluid or solid stress jump.
Here, we use a BJS condition that includes solid deformation velocity in the interfacial velocity jump for its generality~\cite{badia2009coupling,ruiz2022biot,Finney2025}.

The objective of this work is to quantify fluid flow and solid deformation due to transverse and longitudinal peristaltic pumping near a linear poroelastic medium.
We extend the model of \citet{Trevino2025} to identify the flow and deformation dependence on poroelastic material properties (i.e., permeability, porosity, slip coefficient, and shear modulus) and the forward flow and reflux conditions. 
The paper is outlined as follows.
In~\cref{sec:model}, we detail our mathematical model, present the dimensionless governing equations, and define the relevant parameters. 
We provide general forms for the solid deformation, Lagrangian drift velocities, and averaged flow rate.
In~\cref{sec:results}, we quantify interface deformation over a range of permeability and slip coefficient. 
We relate the fluid flow in Stokes channel and poroelastic space to the deformation of the elastic matrix. 
By comparing flow quantities to the rigid impermeable limit, we identify the conditions and locations for flow extrema. 
We discuss the implications of the results in~\cref{sec:discussion} and conclude and offer future extensions in~\cref{sec:conclusions}.

\section{Problem setup\label{sec:model}}
\subsection{Geometry and governing equations}
\begin{figure}[t]
    \centering
    \includegraphics[width=.8\linewidth]{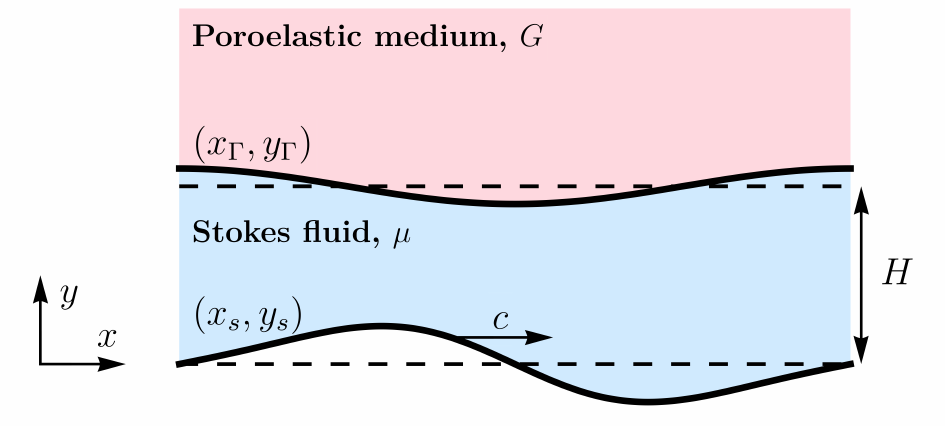}
    \caption{A wave of small amplitude and long wavelength travels in the positive $x$-direction along the bottom boundary. 
    A fluid layer of average thickness $H$ has viscosity $\mu$. 
    A linear poroelastic half space is located at $y>H$ with elastic shear modulus $G$, porosity $\phi$, and permeability $\kappa$.
    Driving wave: $(x_s,y_s)$ and elastic interface: $(x_\Gamma,y_\Gamma)$. 
    Diagram not to scale.}
    \label{fig:schematic}
\end{figure}

\Cref{fig:schematic} shows the problem geometry.
An incompressible fluid layer of viscosity $\mu$ and density $\rho$ is confined between an undulating peristaltic boundary at $y=0$ and an incompressible poroelastic half space at $y>H$. 
The Reynolds number of the flow is assumed to be small.
Thus, the external fluid is governed by the Stokes equations:
\begin{subequations}
\label{eq:stokes}
    \begin{equation}
        \nabla\cdot\mathbf{\Sigma} = \mathbf{0},
    \end{equation}
    \begin{equation}
        \nabla\cdot\mathbf{V} = \mathbf{0},
    \end{equation}
\end{subequations}
where $\mathbf{V}$ is the fluid velocity and $\mathbf{\Sigma}$ the Cauchy stress tensor for a viscous Newtonian fluid given by
    \begin{equation}
        \mathbf{\Sigma} = -P\mathbf{I}+\mu \left[\nabla \mathbf{V} + (\nabla \mathbf{V})^\text{T} \right],
    \end{equation}
and $P$ the fluid pressure.

The region $y>H$ is a fully saturated poroelastic half space composed of an elastic solid skeleton and interstitial viscous fluid identical to the external fluid. 
The fluid and solid phases of the poroelastic medium are incompressible. 
The deformation of the poroelastic medium is also assumed to be infinitesimally small such that the solid is governed by linear elasticity:
\begin{subequations}
\label{eq:cauchy}
    \begin{equation}
        \nabla\cdot\boldsymbol{\sigma} = \mathbf{0},
    \end{equation}
    \begin{equation}
        \nabla\cdot\mathbf{u} = \mathbf{0},
    \end{equation}
\end{subequations}
where $\mathbf{u}$ is the solid deformation field and $\boldsymbol{\sigma}$ the Cauchy stress tensor of the elastic solid given by
    \begin{equation}
        \boldsymbol{\sigma} = -p\mathbf{I}+G \left[\nabla \mathbf{u} + (\nabla \mathbf{u})^\text{T} \right],
    \end{equation}
where $p$ is the pore pressure within the poroelastic region and $G$ the elastic shear modulus. The motion in the poroelastic region is assumed to be dominated by elastic stresses such that viscous shear can be neglected and interstitial flow through the elastic skeleton is given by Darcy's law: 
\begin{equation}
    \nabla{p} = \frac{\phi\mu}{\kappa(\phi)}\left(\frac{D\mathbf{u}}{D t}-\mathbf{v}\right),
    \label{eq:darcy}
\end{equation}
where $\phi\in[0,1]$ is the porosity, i.e., the volume fraction of fluid in the porous region, $\kappa(\phi)$ the permeability, $D/Dt$ the substantial material time derivative, and $\mathbf{v}$ the velocity vector field of the interstitial fluid. 
Increasing permeability reduces the elastic skeleton's resistance to fluid flow such that the material is impervious as $\kappa\rightarrow0$ and has no interstitial flow resistance as $\kappa\rightarrow\infty$.
In general, the porosity and permeability vary in space due to the solid skeleton deformation. 
However, given the deformation is assumed to be infinitesimally small, the porosity and permeability are considered to be constant.

The peristaltic wave is located at coordinates $\mathbf{x}_s = [x_s,y_s]^\text{T}$ and travels in a motion prescribed by
\begin{equation}
    \mathbf{x}_s = \left[\begin{array}{c}
             x+b \cos (k\xi) \\
             d \cos (k\xi)
        \end{array}\right],
\end{equation}
in the positive $x$-direction, where $b$ and $d$ are the longitudinal and transverse peristaltic amplitudes and $k$ the wave number. Here, we define $\xi = (x - c t)$ for wave speed $c$ such that ${\partial}/{\partial \xi}={\partial}/{\partial x}=-{\partial}/{\partial t}$ in subsequent equations.

The Stokes fluid-poroelastic solid interface is located at $\mathbf{x}_{\Gamma} =  \left[x_{\Gamma},y_{\Gamma} \right]^\text{T}$ with coordinates
\begin{equation}
   \mathbf{x}_{\Gamma} = \left[\begin{array}{c}
             x+ u_x(x,H) \\
             H+u_y(x,H)
        \end{array}\right],
\end{equation}
where $u_x$ and $u_y$ are the horizontal and vertical components of the deformation, respectively.

\subsection{Boundary conditions}
The domain is periodic in the $x$-direction. 
At the bottom of the domain, the no-slip and no-penetration boundary conditions between the waving sheet and the fluid yields
\begin{equation}
    \mathbf{V}(\mathbf{x}_s,\xi) = -
    \frac{\partial{\mathbf{x}_s}}{\partial{\xi}} = 
    \left[\begin{array}{c} 
    bk \sin (k\xi) \\ 
    dk \sin(k\xi)  
    \end{array}\right].
    \label{eq:bc1}
\end{equation}
At $\mathbf{x}_\Gamma$, mass continuity and normal and tangential stress balance require
\begin{subequations}
\label{eq:interface}
    \begin{equation}
        (\mathbf{V} - \mathbf{q})\cdot \hat{\mathbf{n}}=0,
    \end{equation}
    \begin{equation}
       \mathbf{\hat{n}} \cdot (\mathbf{\Sigma} - \boldsymbol{\sigma}) \cdot \mathbf{\hat{n}} = 0,
    \end{equation}
    \begin{equation}
        \mathbf{\hat{t}}\cdot\left(\mathbf{\Sigma} - \boldsymbol{\sigma}\right)\cdot \mathbf{\hat{n}}  = 0,
    \end{equation}
\end{subequations}
where $\mathbf{\hat{n}}$ and $\mathbf{\hat{t}}$ are the normal and tangential unit vectors along the poroelastic interface, respectively. 
$\mathbf{q}$ is the total internal velocity of the poroelastic medium:
\begin{equation}
    \mathbf{q} = \phi\mathbf{v}-(1-\phi)\frac{D\mathbf{u}}{D\xi},
    \label{eq:poroelasticVelocity}
\end{equation}
To close the equations, we consider the Beavers-Joseph-Saffman (BJS) slip boundary condition~\cite{Beavers1967,Saffman1971}, i.e.,
\begin{equation}
    \hat{\mathbf{t}}\cdot\frac{\partial\mathbf{V}}{\partial \hat{\mathbf{n}}}=\frac{\gamma}{\sqrt{\kappa}}(\mathbf{V} - \mathbf{q})\cdot\hat{\mathbf{t}} ,
    \label{eq:bjs}
\end{equation}
where $\gamma$ is the dimensionless empirical interfacial slip, hereby referred to as slip.
The boundary condition is no-slip as $\gamma\rightarrow\infty$ and perfect slip as $\gamma=0$.
The poroelastic solid deformation decays vertically:
\begin{equation}
    \mathbf{u}(\xi,y\rightarrow\infty)=\mathbf{0}.
    \label{eq:decay}
\end{equation}

\subsection{Solution}
The governing equations are non-dimensionalized using the respective characteristic length and time scales $1/k$ and $1/ck$. 
The dimensionless parameters are $\Tilde{H}=Hk$, $\Tilde{b}=bk$,     $\Tilde{d}=dk$, $\Tilde{t}=tck$, $\Tilde{x}=xk$, $\Tilde{\kappa}=\kappa k^2$, $\Tilde{P} = P /(\mu ck)$, and $\mathit{\Lambda} = G/{(\mu ck)}$.
The stiffness parameter $\mathit{\Lambda}$ is the ratio of elastic to viscous interfacial stress such that large and small $\mathit{\Lambda}$ represents a stiff and highly deformable material, respectively. 
It is assumed the elastic forces dominate such that $\mathit{\Lambda}\geq1$.
In general, the permeability, slip, and stiffness vary across orders of magnitude depending on material and system scale~\citep{Trevino2025,Finney2025}.
We choose parameter ranges $\tilde\kappa\in[10^{-4},10]$, $\tilde\gamma\in[0,10^4]$, and $\mathit{\Lambda}\in[1,10^4]$ to encompass the full behavior of the field quantities.
For simplicity, tildes are subsequently dropped.
The dimensionless governing equations in the fluid and poroelastic solid are 
\begin{subequations}
\label{eq:stokes_dimless}
    \begin{equation}
        \nabla P = \nabla^2\mathbf{V},
    \end{equation}
    \begin{equation}
        \nabla\cdot\mathbf{V} = \mathbf{0},
    \end{equation}
\end{subequations}
and
\begin{subequations}
\label{eq:cauchy_dimless}
    \begin{equation}
        \nabla p = \mathit{\Lambda}\nabla^2\mathbf{u},
    \end{equation}
    \begin{equation}
        \nabla\cdot\mathbf{u} = \mathbf{0},
    \end{equation}
\end{subequations}
respectively. 
The dimensionless Darcy's law is
\begin{equation}
\label{eq:darcydimensionless}
    \mathbf{v} = \frac{D\mathbf{u}}{D \xi}-\frac{\kappa}{\phi}\nabla p.
\end{equation}
In this rearranged form, it is evident that the second term on the right hand side becomes non-negligible as permeability increases. 
Moreover, this term represents the influence of the solid skeleton on the interstitial fluid being reduced as the resistance to flow is reduced.
The no-slip boundary condition along the peristaltic wave becomes
\begin{equation}
    \mathbf{V}(\mathbf{x}_s,t) = -
    \frac{\partial{\mathbf{x}_s}}{\partial{\xi}} = 
    \left[\begin{array}{c} 
    b \sin (\xi) \\ 
    d \sin(\xi)  
    \end{array}\right],
    \label{eq:y0dimless}
\end{equation}
and \cref{eq:interface,eq:bjs} at the poroelastic interface $\mathbf{x}_\Gamma$ are
\begin{subequations}
\label{eq:bc}
    \begin{equation}
        V_y - \left(\phi v_{y}-(1-\phi)\frac{D u_y}{D\xi}\right) = 0,
    \end{equation}
    \begin{equation}
        \label{eq:bjs1}
        \frac{\partial V_x}{\partial y}-\frac{\gamma}{\sqrt{\kappa}}\left(V_x+\frac{D u_x}{D \xi}\right) = 0,
    \end{equation}    
    \begin{equation}
     \label{eq:norm}
       p +2\mathit{\Lambda}\frac{\partial u_y}{\partial y} - \left( P + 2\frac{\partial V_y}{\partial y}\right) = 0,     
    \end{equation}
    \begin{equation}
         \label{eq:shear}
        \mathit{\Lambda} \left(\frac{\partial u_x}{\partial y}+\frac{\partial u_y}{\partial x}\right) - \left(\frac{\partial V_x}{\partial y}+\frac{\partial V_y}{\partial x}\right) = 0.
    \end{equation}
\end{subequations}
We note that the Stokes drift in \cref{eq:bjs1} decreases by either decreasing $\gamma$ or increasing $\kappa$. 
With the incompressibility of the Stokes fluid and elastic solid, we define the stream functions $\Psi$ and  $\psi$, such that $\mathbf{V}=\nabla \times \Psi\hat{\mathbf{z}}$ and $\mathbf{u}=\nabla \times \psi\hat{\mathbf{z}}$. 
\Cref{eq:stokes_dimless,eq:cauchy_dimless} imply $\nabla^4\Psi=0$ and $\nabla^4\psi=0$, respectively. 
The general solutions to the stream functions that satisfy the problem geometry are 
\begin{subequations}    
\label{eq:stream}
    \begin{equation}
    \begin{aligned}
    \Psi^{(n)}= & \sum_{m=1}^{\infty}\left\{\left[\left(A_{ m}^{(n)}+E_{m}^{(n)} {y}\right) \cos m\xi\right.\right.  \left.+\left(B_{m}^{(n)}+F_{m}^{(n)} {y}\right) \sin m\xi\right] \cosh m {y} \\
    & +\left[\left(C_{ m}^{(n)}+G_{ m}^{(n)} {y}\right) \cos m\xi\right. \left.\left.+\left(D_{m}^{(n)}+H_{m}^{(n)} {y}\right) \sin m(\xi)\right] \sinh m {y}\right\} \\
    & +\alpha_n {y}+\beta_n {y}^2+\gamma_n {y}^3,
    \end{aligned}
    \end{equation}
    \begin{equation}
    \begin{aligned}
        \psi^{(n)}=  &\sum_{m=1}^{\infty}\left[\left(A_{ m}^{(n)}+E_{m}^{(n)}{y}\right) \cos m\xi\right.  \left.+\left(B_{ m}^{(n)}+F_{m}^{(n)} {y}\right) \sin m\xi\right] e^{-m {y}},
    \end{aligned}
\end{equation}
\end{subequations}
for order $n$ and mode $m$.
To solve the system of equations, we expand \cref{eq:y0dimless} about $y=0$ and \cref{eq:bc} about $y=H$ in orders of the small peristaltic amplitude, $dk,~bk\ll1$, respectively. 
The coefficients of \cref{eq:stream} are then solved order by order for $\mathbf{V}$ and $\mathbf{u}$~\cite{Trevino2025}. 
The velocity of the interstitial fluid is solved using \cref{eq:darcydimensionless}.

\subsection{Field quantities}
Elastic matrix deformation and Lagrangian drift velocities are calculated at first order in the peristaltic amplitude and Eulerian flow rate at second order.
The first order oscillatory Stokes flow drives poroelastic oscillatory deformations of the form
\begin{equation}
    \mathbf{u} = \left[\begin{array}{c}
    \mathcal{A}\cos\xi+\mathcal{B}\sin\xi \\
    \mathcal{C}\cos\xi+\mathcal{D}\sin\xi
    \end{array}\right],
    \label{eq:abcd}
\end{equation}
where the wave amplitudes $\mathcal{A}$, $\mathcal{B}$, $\mathcal{C}$, and $\mathcal{D}$ are functions of permeability, stiffness, and slip. 
The deformations are determined using \cref{eq:bjs1,eq:norm,eq:shear} and normalized with the persitaltic amplitude. 
The undulating boundary motion induces horizontal Lagrangian drift of the Stokes and interstitial pore fluid, found respectively by
\begin{subequations}
    \begin{equation}
    V_{\text{Lag}}(y)=\frac{1}{2\pi}\int \left( \frac{\partial V_x^{(1)}}{\partial x}\delta x + \frac{\partial V_x^{(1)}}{\partial y}\delta y\right) ~d\xi,
    \label{eq:vlag}
\end{equation}
    \begin{equation}
    v_{\text{Lag}}(y)=\frac{1}{2\pi}\int \left( \frac{\partial v_{x}^{(1)}}{\partial x}\delta x + \frac{\partial v_{x}^{(1)}}{\partial y}\delta y\right) ~d\xi.
    \label{eq:vflag}
\end{equation}
\end{subequations}
Differential particle trajectories $\delta x$ and $\delta y$ of a given point are obtained by integrating the first order velocity \cite{Lighthill_1992,Andrews_Mcintyre_1978,ibanez2021simple}.
We shall refer to $V_{\text{Lag}}$ and $v_{\text{Lag}}$ as the Stokes and Darcy drift, respectively, and $v_x$ as the Darcy velocity.
$V_x^{(1)}$ and $v_{x}^{(1)}$ are the respective $x$-components of the first order Stokes and Darcy flow.
The Lagrangian drift is the average horizontal motion of a fluid parcel over a wavelength of the peristaltic wave.
The average Eulerian flow rate through the Stokes channel is calculated with:
\begin{equation}
    Q  =  \frac{1}{2 \pi}\int_{0}^{2 \pi}\int_{0}^{H} V_x\,dy \,d\xi,
\end{equation}
where $V_x$ is the second order horizontal Stokes velocity.
We define the flow velocity at the peristaltic wave in impermeable rigid confinement as a normalization factor:
\begin{equation}
    \begin{aligned}
V_0=\lim_{\substack{\{\kappa,\phi\} \to 0\\
\{\gamma,\Lambda\} \to\infty}} V_x|_{y=0}.
    \end{aligned}
\end{equation}
The flow rate is normalized by the flow rate near a rigid impermeable boundary~\citep{Trevino2025}:
\begin{equation}
    \begin{aligned}
Q_R=\lim_{\substack{\{\kappa,\phi\} \to 0\\
\{\gamma,\Lambda\} \to\infty}} Q    =\frac{1}{4}H\left[-b^2+ d^2 \left(1+\frac{4 H^2}{-2 H^2+\cosh (2 H)-1}\right)\right].
    \end{aligned}
\end{equation}

\section{Results\label{sec:results}}
\subsection{Deformation}
In the limit of $\kappa\rightarrow0,~\gamma\rightarrow\infty$, and $\phi=0$, the poroelastic interface deformation solution matches the impermeable linear elastic result in \citet{Trevino2025}. 
Specifically, the solid deformation monotonically decreases with increasing material stiffness. 
Further, deformation is negligibly dependent on porosity and so we set $\phi=0.5$, $\mathit{\Lambda}=1$, and $H =0.5$.
Moreover, permeability primarily acts normal to the interface and negligibly affects $u_x$. 
Similarly, slip primarily acts tangent to the interface and minimally influences $u_y$.

\begin{figure}[t]
    \centering
    \includegraphics[width=\linewidth]{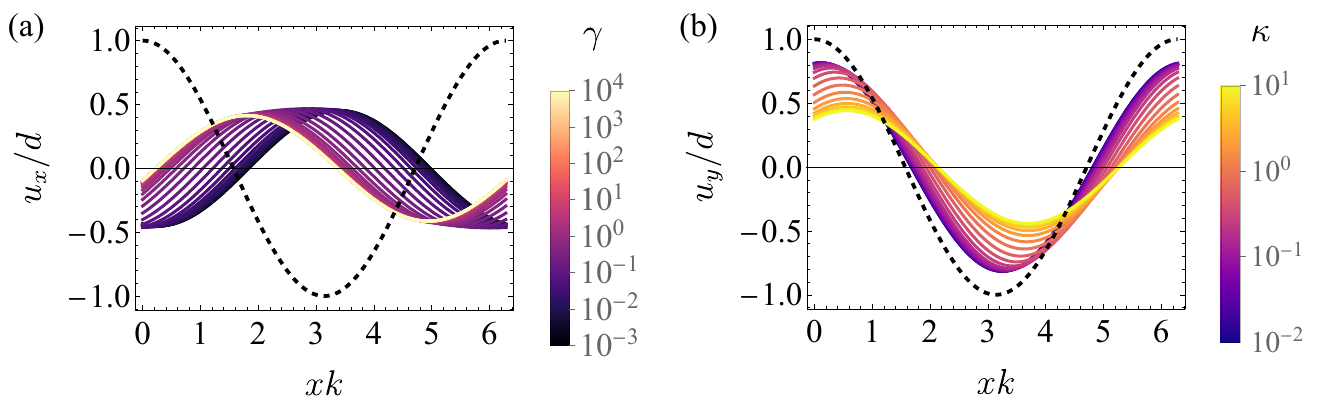}
    \caption{Transverse peristalsis deformation of the poroelastic interface vs $xk$-direction. 
    (a): horizontal, slip varies and $\kappa=0.01$ and (b): vertical,  permeability varies and $\gamma=100$.
    Dashed black line: the driving wave.} 
    \label{fig:deform_kg}
\end{figure}

\begin{figure}[t]
    \centering
    \includegraphics[width=\linewidth]{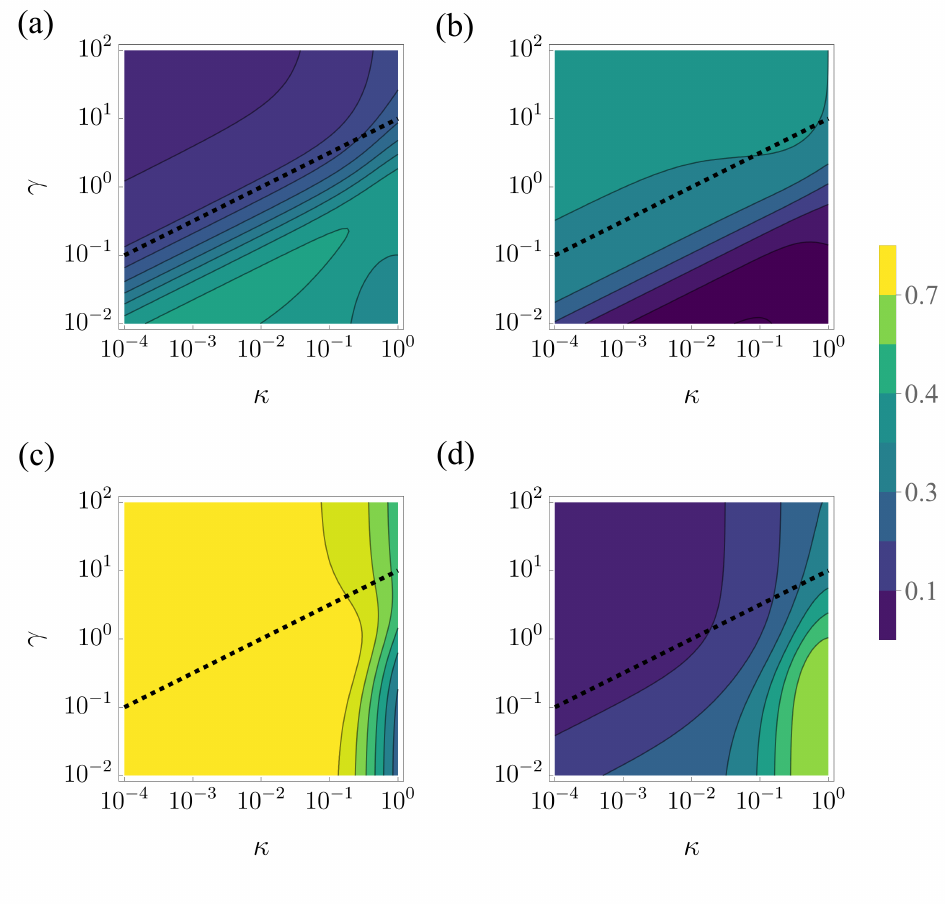}
    \caption{Contours of transverse peristalsis amplitudes vs permeability and interfacial slip.
    (a): $|\mathcal{A}|/d$, (b):  $|\mathcal{B}|/d$, (c): $|\mathcal{C}|/d$, and (d): $|\mathcal{D}|/d$.} 
    \label{fig:abcd_trans}
\end{figure}

\Cref{fig:deform_kg,fig:abcd_trans} show the respective surface deformation and contours of the absolute wave amplitudes as functions of permeability and slip for transverse peristalsis.
Slip monotonically changes the shape of $u_x$ from  sinusoidal at large $\gamma$ to co-sinusoidal as $\gamma$ decreases. 
Increasing permeability decreases the magnitude and shifts the phase of $u_y$ by a quarter of a period. 
As $\gamma$ increases from $\SI{E-2}{}$ to $\SI{E2}{}$, the interface deformation transitions from co-sinusoidal to sinusoidal as $|\mathcal{A}|$ becomes smaller than $|\mathcal{B}|$. 
$|\mathcal{A}|$ and $|\mathcal{B}|$ are approximately constant along lines of constant $\gamma/\sqrt{\kappa}$, arising from the coefficient on the right hand side of \cref{eq:bjs1}.
The vertical displacement remains in phase with the peristaltic wave since $|\mathcal{C}| = 0.7$ for $\kappa<10^{-1}$ and $|\mathcal{D}|<0.3$ except for $\kappa>10^{-1}$ and $\gamma<1$.


\begin{figure}[t]
    \centering
    \includegraphics[width=\linewidth]{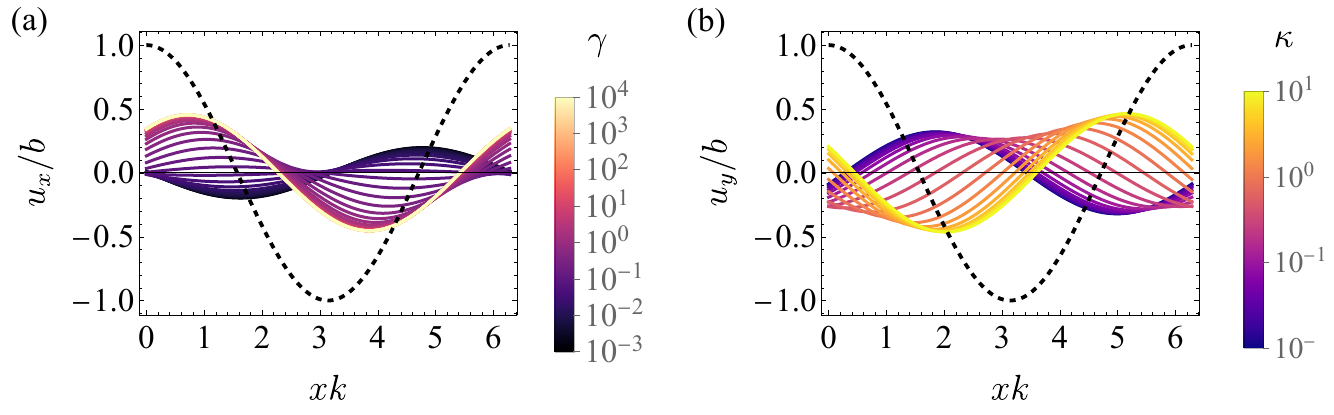}
    \caption{Longitudinal peristalsis deformation of the poroelastic interface vs $xk$-direction. 
    (a): horizontal, slip varies and $\kappa=0.01$ and (b): vertical,  permeability varies and $\gamma=100$.
    Dashed black line: the driving wave.} 
    \label{fig:deform_kg_long}
\end{figure}

\begin{figure}[t]
    \centering
    \includegraphics[width=\linewidth]{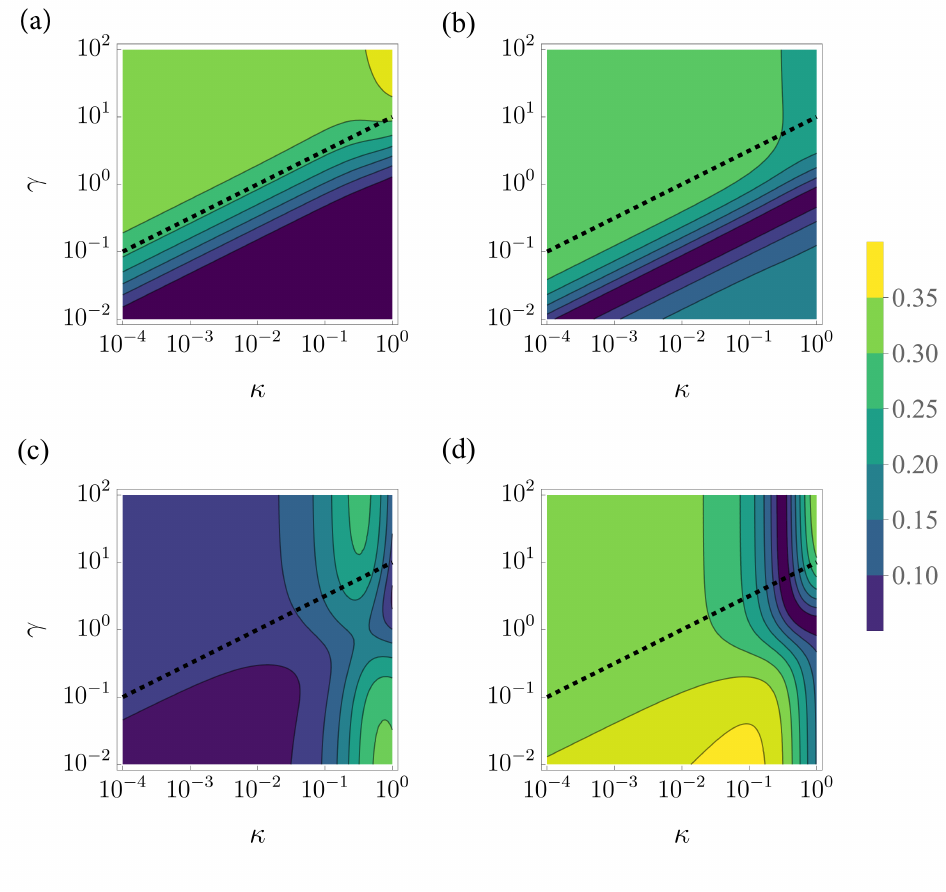}
    \caption{Contours of longitudinal peristalsis amplitudes (a): $|\mathcal{A}|/b$, (b): $|\mathcal{B}|/b$, (c): $|\mathcal{C}|/b$, and (d): $|\mathcal{D}|/b$ vs permeability and interfacial slip.} 
    \label{fig:abcd_long}
\end{figure}

\Cref{fig:deform_kg_long,fig:abcd_long} show the respective surface deformation and contours of the absolute wave amplitudes as functions of permeability and slip for longitudinal peristalsis.
The deformation magnitudes are relatively smaller than those from transverse waves.
As the slip increases from $\SI{E-3}{}$ to $\SI{E4}{}$, magnitude of $u_x$ increases and shifts out of phase by almost a wave period. 
On the other hand, increasing the permeability shifts the phase of $u_y$ by half a wave period.
For $\gamma<\SI{1}{}$ and $\kappa < \SI{E-1}{}$, $|\mathcal{A}|<0.1$ and $|\mathcal{B}|$ is  proportional with $\gamma/\sqrt{\kappa}$. 
For $\kappa<\SI{E-1}{}$, $|\mathcal{D}|>|\mathcal{C}|$ and $u_y$ is sinusoidal. 


\subsection{Lagrangian mean velocities}
Porosity acts within the bulk of porous region and has a negligible effect on $V_{\text{Lag}}$.
Thus, for the remainder of the paper, porosity and channel thickness are held constant at $\phi=0.5$ and $H=1$, respectively. 
In the transverse case, the Stokes and Darcy drifts are scaled by $10^2$ and $10^3$, respectively. In the longitudinal case, the Stokes and Darcy drifts are scaled by $10^2$ and $10^3$, respectively. 

\begin{figure}[t]
    \centering
    \includegraphics[width=\linewidth]{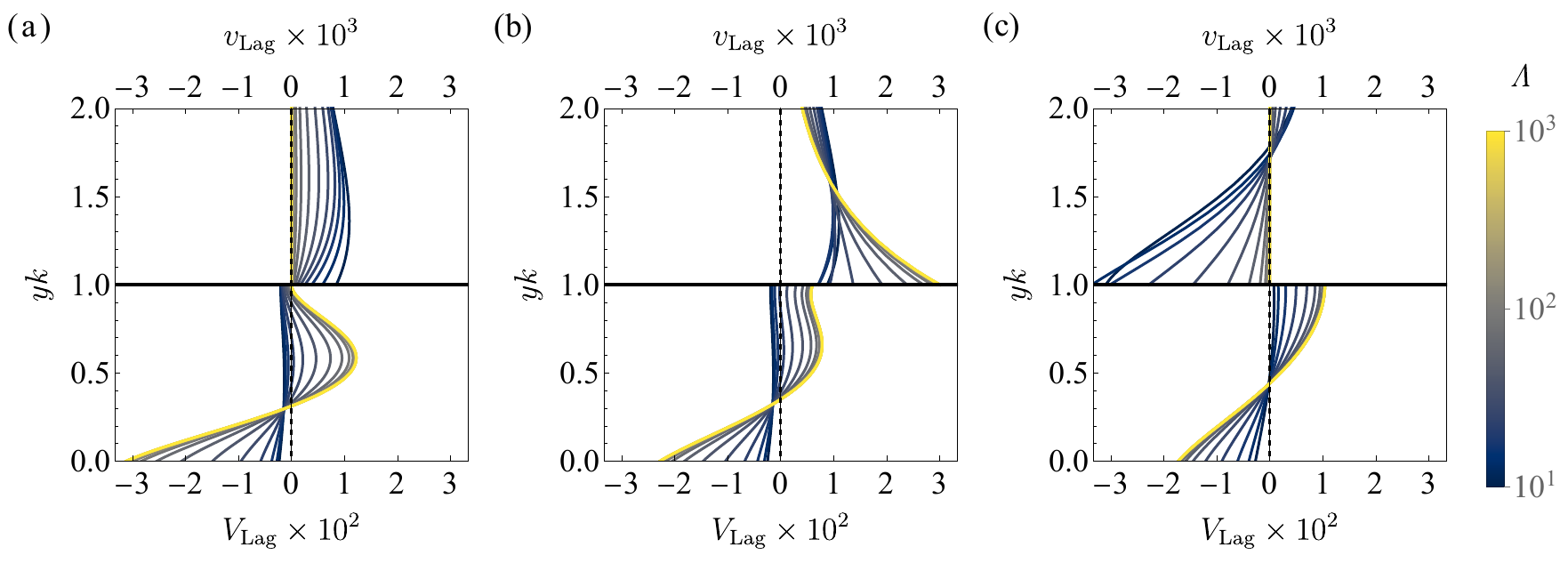}
    \caption{Darcy (top) and Stokes (bottom) drifts for transverse peristalsis over a range of different stiffness parameters. 
    (a): $\kappa=10^{-4}$ and $\gamma=10^4$. 
    (b): $\kappa=10^{-1}$ and $\gamma=10^4$. 
    (c): $\kappa=10^{-4}$ and $\gamma=0$. 
    Horizontal black line: unperturbed poroelastic interface.} 
    \label{fig:v_lag_trans}
\end{figure}

\Cref{fig:v_lag_trans} shows transverse peristaltic Darcy (top) and Stokes (bottom) drift velocity profiles for varying stiffness at three combinations of permeability and slip.  
For low permeability ($\kappa=10^{-4}$) and no-slip ($\gamma=10^4$)  [\Cref{fig:v_lag_trans}(a)], the Stokes drift profiles matches those of the impermeable elastic case (Figure 8(a) in \citet{Trevino2025}). 
Reflux occurs near the transverse peristaltic wave and forward flow near the upper boundary~\citep{ibanez2021simple,Lighthill_1992,Andrews_Mcintyre_1978,Shapiro1969}. 
As the boundary becomes more compliant, the velocity extrema reduce, implying a loss in kinetic energy due to decreasing stiffness.
Conversely, the Darcy drift monotonically increases as the material softens due to elastic matrix motion (see \cref{eq:darcydimensionless}). 
Maintaining no-slip and increasing permeability [\cref{fig:v_lag_trans}(b)], a nonzero drift velocity is present as $\mathit{\Lambda}\rightarrow\infty$ at the boundary in both flow regions.
The oscillatory fluid motion within the Stokes channel drives the Darcy flow rather than the elastic skeleton motion.
Specifically, Stokes flow at the interface loads the pore pressure, resulting in a mean drift.
In this limit, increasing solid softness decreases the Darcy drift.
For a perfect slip condition ($\gamma=0$) with small permeability [\cref{fig:v_lag_trans}(c)], an expected nonzero Stokes drift at the interface is observed at high stiffness (see \cref{eq:bjs1}).
However, Darcy drift is zero at the poroelastic interface, indicating a large permeability and $\mathit{\Lambda}$ is necessary to produce a non-zero Darcy drift. 

\begin{figure}[t]
    \centering
    \includegraphics[width=\linewidth]{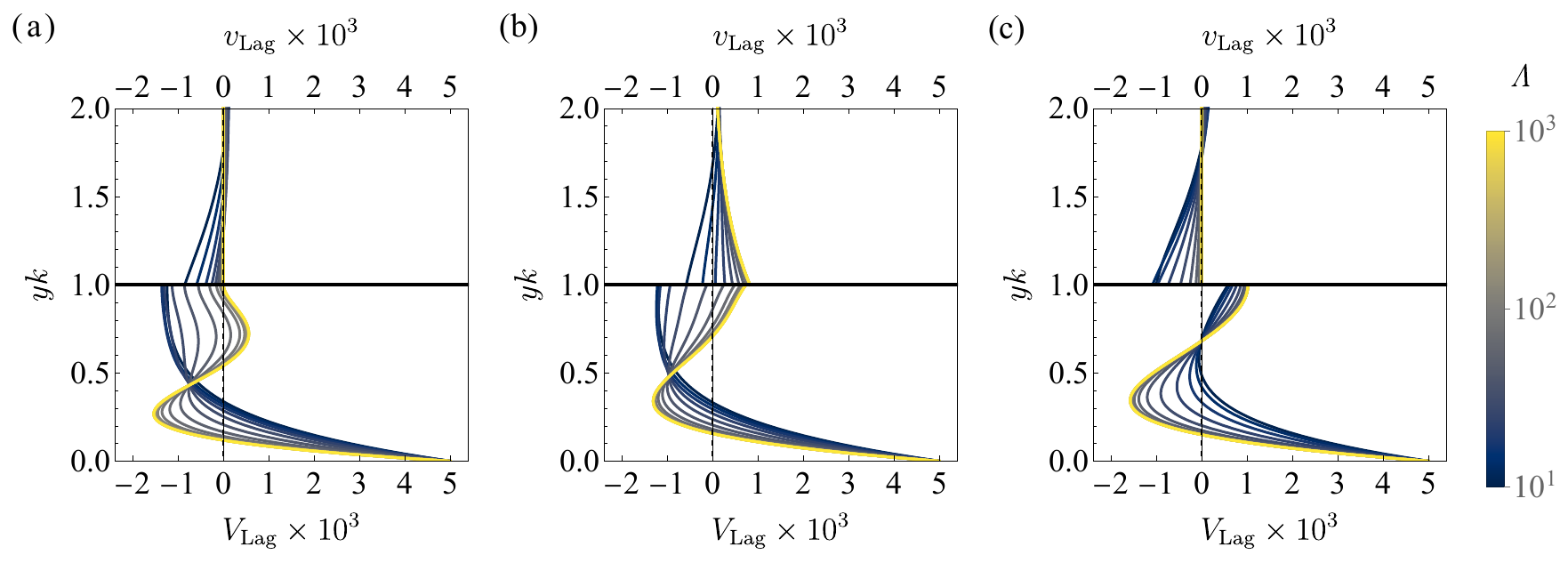}
    \caption{Darcy (top) and Stokes (bottom) drifts for longitudinal peristalsis over a range of stiffness parameters. 
    (a): $\kappa=10^{-4}$ and $\gamma=10^4$. 
    (b): $\kappa=10^{-1}$ and $\gamma=10^4$. 
    (c): $\kappa=10^{-4}$ and $\gamma=0$. 
    Horizontal black line: unperturbed poroelastic interface.}
    \label{fig:v_lag_long}
\end{figure}

\Cref{fig:v_lag_long} shows the Stokes and Darcy drift for longitudinal peristaltic wave motion across a range of stiffness parameter values for three combinations of slip and permeability.
Unlike transverse peristalsis, Stokes drift near the driving wave are positive in longitudinal peristalsis. 
Thus, coupling longitudinal and transverse peristaltic motion can reduce fluid reflux~\citep{Kalayeh2023,Trevino2025}.
Aside from the reversed flow direction, the Stokes and Darcy drifts trends are the same between longitudinal and transverse peristalsis.

\subsection{Eulerian flow}
\begin{figure}[t]
    \centering
    \includegraphics[width=\linewidth]{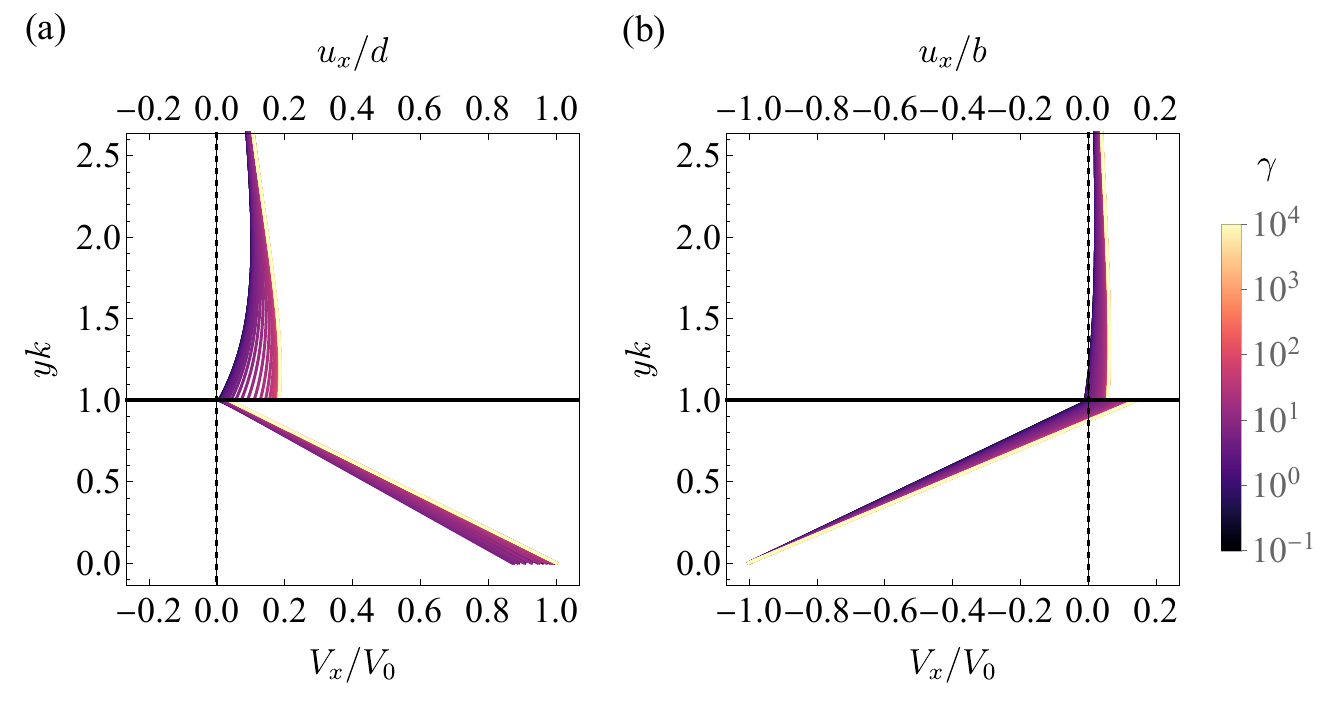}
    \caption{Porous medium deformation (top) and Stokes velocity (bottom) for transverse (a) and longitudinal (b) peristalsis for different interfacial slips.
    Horizontal black line: interface.}
    \label{fig:v_prof_gam}
\end{figure}

\Cref{fig:v_prof_gam} shows the $xk$-direction streaming Stokes velocity (bottom) and the first order porous medium deformation (top) as functions of $yk$ for transverse (left) and longitudinal (right) peristalsis.
The time-averaged second order Stokes velocity profiles reduce to simple shear flow driven by the peristaltic wave.
The transverse and longitudinal wave produce net forward Eulerian flow and reflux \citep{Kalayeh2023,Trevino2025}, respectively.
As expected, the horizontal solid deformation is maximized in the no-slip limit ($\gamma\rightarrow\infty$).
Similarly, as $\gamma\rightarrow0$, the horizontal deformation at the interface approaches zero.

\begin{figure}[t]
    \centering
    \includegraphics[width=\linewidth]{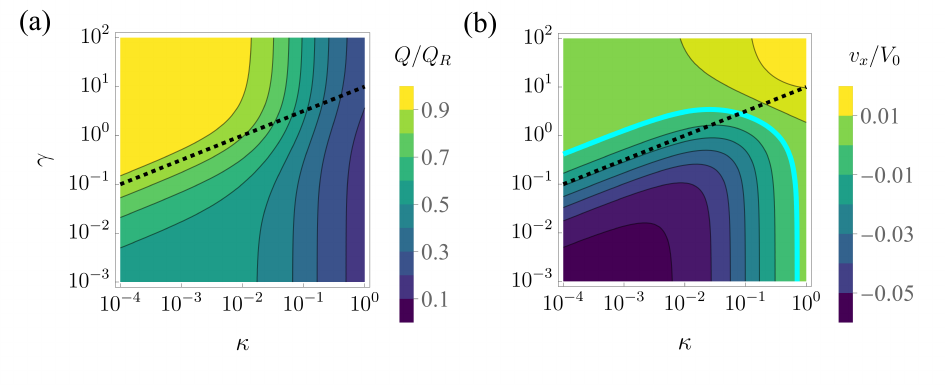}
    \caption{Contours of the (a) flow rate within the Stokes fluid channel and (b) Darcy velocity under transverse peristaltic pumping.
    Black dashed line: $\gamma/\sqrt{\kappa}=10$,
    cyan line in (b): $v_x/V_0=0$.}
    \label{fig:contour_Q_trans}
\end{figure}

\Cref{fig:contour_Q_trans} shows contours of the normalized flow rate and the Darcy velocity for a range of $\kappa$ and $\gamma$ induced by transverse peristaltic motion. 
The flow rate is independent of permeability and slip when $\mathit{\Lambda}\lesssim\mathcal{O}(10)$ in the Stokes fluid channel (data not shown) and $\mathit{\Lambda}\gtrsim\mathcal{O}(100)$.
Thus, we set $\mathit{\Lambda}=10^4$ for \cref{fig:contour_Q_trans}(a) and $\mathit{\Lambda}=10$ for \cref{fig:contour_Q_trans}(b), to show the permeability and slip dependence.
The flow rate through the Stokes channel is maximized for $\gamma>\SI{E-1}{}, \kappa<\SI{E-2}{}$.
As the upper region becomes fluid-like ($\gamma\rightarrow0$, $\kappa\rightarrow\infty$), the rigid confinement effect on the flow is reduced. 
The flow rate approaches zero as the interfacial slip goes to zero and permeability to unity.
Contrary to the positive flow within the Stokes channel, the Darcy velocity undergoes reflux when either $\gamma$ or  $\kappa<1$ and produces forward flow otherwise. 
Moreover, it reaches its maximum magnitude during reflux and increases along diagonal contours of constant $\gamma/\sqrt{\kappa}$.

\begin{figure}[t]
    \centering
    \includegraphics[width=\linewidth]{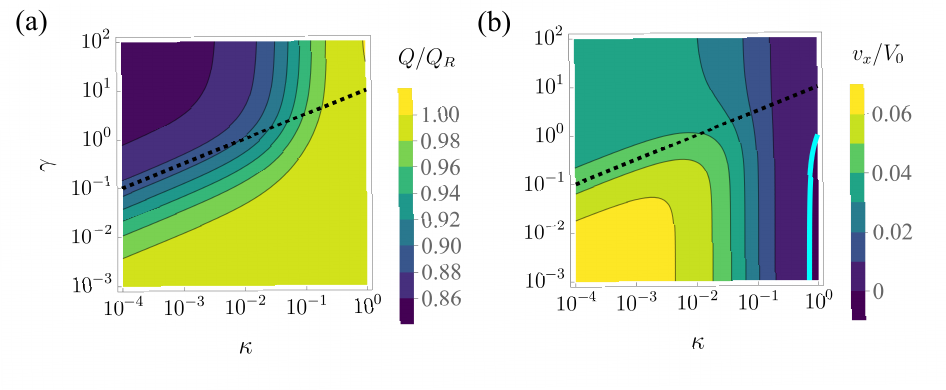}
    \caption{Contours of the (a) flow rate within the Stokes fluid channel and (b) Darcy velocity under longitudinal peristaltic pumping.
    Black dashed line: $\gamma/\sqrt{\kappa}=10$,
    cyan line in (b): $v_x/V_0=0$.}
    \label{fig:contour_Q_long}
\end{figure}

\Cref{fig:contour_Q_long} shows contours of the Stokes channel flow rate and Darcy velocity under longitudinal peristalsis. 
Both the Stokes fluid and Darcy flow undergo reflux and $\mathit{\Lambda}=10$ to highlight the permeability and slip dependence.
Since $Q_R<0$ and $V_0<0$ for longitudinal peristalsis, the contours show positive normalized Stokes fluid and Darcy flow.
The flow rate from the rigid impermeable boundary limit is proportional to permeability and inversely proportional to slip.
The Darcy flow is generally inversely proportional to permeability and along lines of constant $\gamma/\sqrt{\kappa}$.

\section{Discussion\label{sec:discussion}}
We investigated the effect of poroelastic material stiffness, permeability, and slip on a confined peristaltic Stokes flow. 
Since the slip determines the drag tangent to the poroelastic interface  (\cref{eq:bjs1}), it is expected and observed that the interfacial horizontal deformation dominantly depends on slip compared to permeability for both transverse and longitudinal peristalsis.
Vertical deformation is inversely proportional to permeability and changes negligibly with slip.
As permeability increases, the elastic matrix resists fluid transmission across the interface less and therefore deforms less.
Poroelasticity of the boundary decreases the flow rate in the Stokes fluid channel. 
The Stokes flow loses kinetic energy to elastic storage in the solid skeleton and to interstitial viscous dissipation.
On the other hand, it is found that decreasing material rigidity
generally increases the Darcy flow because the motion of the
solid serves to pump interstitial fluid.
This effect diminishes for high permeability where the skeleton-interstitial fluid interaction becomes negligible.
This is reflected in the magnitude of the Darcy flow decreasing for $\kappa>10^{-2}$ and small $\gamma$ (see \cref{fig:contour_Q_trans}(b) and \cref{fig:contour_Q_long}(b)).
This indicates that at a certain $\kappa$, the
reduction in drag past the solid skeleton is no longer advantageous as viscous dissipation
through the pores becomes dominant.

The behavior of our field quantities closely match those of a poroelastic particle in a shear and pressure driven flow reported in \citet{Finney2025}. 
The authors vary the Poisson ratio of the poroelastic particle, thereby altering relative transverse and tangential strains over the surface of the particle.
Our transverse and longitudinal peristalsis-driven deformations resemble their maximized ($\nu=1/2$) and minimized ($\nu=0$) transverse strains, respectively (see \cref{fig:abcd_trans}(d), \cref{fig:abcd_long}(d), and Fig. 6 of~\citep{Finney2025}).
Additionally, the behavior of Darcy pressure and surface traction as functions of permeability and slip closely match our Darcy flow observations and are inherently related through \cref{eq:darcydimensionless} (e.g., see Fig.~5 of~\citep{Finney2025}).

Our model provides insights into poroelastic physics and can guide material parameter selection in engineered systems.
As an example, consider cerebrospinal fluid flow through perivascular spaces with the following parameter values from experiments \citep{Thomas2019,kelley2021}: $[\phi,~\kappa,~\mathit{\Lambda,~H,~d}]=[0.2,~10^{-5},~10^5,~10^{-3},~10^{-4}]$.
The slip parameter of brain tissue is unknown, thus we consider a range of $\gamma=[10^{-3},10^2]$.
Calculated Eulerian flow velocities are on the order of $\mathcal{O}(\SI{10}{\micro\meter\per\second})$ and in agreement with their \textit{in vivo} observations.
The velocity increases by $45\%$ proportionally with slip.
The slip value uncertainty presents an opportunity to inversely characterize the material slip from experimental flow measurements with our model. 
To study drug delivery (e.g., spinal injury), some lab-on-a-chip devices are designed to mimic biological tissue interacting with periodic flow~\citep{koroleva2016hydrogel,huh2010reconstituting,kim2012human,kwon2025convective,zilberman2021}.
Our model offers a methodology of identifying the material properties necessary to control the flow for directed delivery or filtration of solutes.
For larger scale systems such as wave-induced flow of fluid through porous seabeds~\citep{habel2005wave,webber2021stokes,weber2023wave} and oscillatory groundwater flows~\citep{trefry2019temporal,meza2022flow}, our model can be used to predict the transport of nutrients or potentially harmful particulate matter through poroelastic media.
Future consideration could focus on solute transport by calculating the Taylor dispersion within the Stokes channel and poroelastic region, informing studies of particulate motion~\citep{gan2025antidispersion}.
Alternative boundary conditions can be incorporated into our model to inform peristaltic pumping experiment observations~\citep{ochoa1995momentum1,ochoa1995momentum2,Minale2014,Minale2014_2,Feng2020,xu}.
Additionally, a viscous Brinkman term can also be introduced to the modified Darcy law to generate a viscous shear layer that enhances the rigid limit Stokes flow \citep{Iqbal2025}.

\section{Conclusions\label{sec:conclusions}}
We propose a model for the Stokes flow of small amplitude peristaltic pumping confined by a linear poroelastic medium.
We distinguish two pumping mechanisms in allowing transverse and longitudinal peristalsis. 
In each case, poroelastic material stiffness, permeability, and slip act in concert to influence the Stokes flow, interstitial Darcy flow, and poroelastic deformation.
In general, the external Stokes flow is reduced for a fluid-like (i.e., small stiffness, high permeability, and small slip) poroelastic boundary, owing to increased energy loss through the interface.
Conversely, the Darcy flow increases with decreasing material stiffness and has a non-uniform dependence on permeability and slip.
It reaches a maximum velocity when permeability is low enough for the solid skeleton to pump interstitial flow without sacrificing energy to viscous dissipation.
Future work will numerically simulate nonlinear poroelastic finite deformations to admit spatio-temporal changes to porosity and permeability. 

\section{Acknowledgments}
The authors thank Thomas R. Powers for fruitful discussions during the preparation of this work. 
MRJ acknowledges support from the U.S. Department of Defense under the DEPSCoR program Award No. FA9550-23-1-0485 (PM Dr. Timothy Bentley) and Brown University Seed Research award.
MRJ and RZ acknowledge the support from the School of Engineering Hazeltine Innovation Award.
Funding agencies were not involved in study design; in the collection, analysis and interpretation of data; in the writing of the report; or in the decision to submit the article for publication.
The opinions, findings, and conclusions, or recommendations expressed are those of the authors and do not necessarily reflect the views of the funding agencies.

\section{Data availability}
The code for this work is available at:

\bibliography{refs}
\end{document}